\documentclass[twoside]{article}
\usepackage{qic,epsfig,braket,amsmath}

\begin{document}
\setlength{\textheight}{8.0truein}

\runninghead{ENTANGLEMENT PURIFICATION WITH TWO-WAY CLASSICAL
COMMUNICATION}{ALAN W. LEUNG AND PETER W. SHOR}

\normalsize\textlineskip
\thispagestyle{empty}
\setcounter{page}{1}


\vspace*{0.88truein}

\alphfootnote

\fpage{1}

\centerline{\bf Entanglement purification with two-way classical
communication}
\vspace*{0.37truein}

\centerline{\footnotesize Alan W. Leung\footnote{E-mail address:
leung@math.mit.edu}} \vspace*{0.015truein}
\centerline{\footnotesize\it Department of Mathematics,
Massachusetts Institute of Technology,} \baselineskip=10pt
\centerline{\footnotesize\it 77 Massachusetts Avenue, Cambridge,
MA 02139, USA} \vspace*{10pt} \centerline{\footnotesize Peter W.
Shor\footnote{E-mail address: shor@math.mit.edu}}
\vspace*{0.015truein} \centerline{\footnotesize\it Department of
Mathematics, Massachusetts Institute of Technology,}
\baselineskip=10pt \centerline{\footnotesize\it 77 Massachusetts
Avenue, Cambridge, MA 02139, USA} \vspace*{0.225truein}

\vspace*{0.21truein}

\abstracts{We present an improved protocol for entanglement
purification of bipartite mixed states. The protocol requires
two-way classical communication and hence implies an improved
lower bound on the quantum capacity with two-way classical
communication of the quantum depolarizing channel.}{}{}

\vspace*{10pt}


\vspace*{1pt}\textlineskip    
\section{Introduction}
Quantum information theory and quantum computation study the use
of quantum physics in information processing and
computation\cite{NC}. Many important results such as quantum
teleportation, superdense coding, factoring and search algorithms
make use of quantum entanglements as fundamental resources
\cite{BBCJPW, BW, S, G}. Therefore, entanglement purification
protocols, the procedures by which we extract pure-state
entanglements from mixed states, merit our study.

In this work, we follow the framework of \cite{BDSW, MS} and
present a new purification protocol with improved yields

\subsection{Notation}\label{1.1}
We denote von Neumann entropy by $S(\rho)$ and Shannon entropy by
$H(p_0,p_1,\ldots)$. The following notation is used for the four
Bell states:

{\setlength\arraycolsep{2pt}
\begin{eqnarray}
\ket{\Phi^{+}}=\frac{1}{\sqrt{2}}(\ket{\uparrow\uparrow}+\ket{\downarrow\downarrow})\nonumber\\
\ket{\Psi^{+}}=\frac{1}{\sqrt{2}}(\ket{\uparrow\downarrow}+\ket{\downarrow\uparrow})\nonumber\\
\ket{\Phi^{-}}=\frac{1}{\sqrt{2}}(\ket{\uparrow\uparrow}-\ket{\downarrow\downarrow})\nonumber\\
\ket{\Psi^{-}}=\frac{1}{\sqrt{2}}(\ket{\uparrow\downarrow}-\ket{\downarrow\uparrow})
\end{eqnarray}}

\noindent To facilitate our discussion later on, we also use two
classical bits to label each of the Bell states:

{\setlength\arraycolsep{2pt}
\begin{eqnarray}
\Phi^{+}=00\nonumber\\
\Psi^{+}=01\nonumber\\
\Phi^{-}=10\nonumber\\
\Psi^{-}=11
\end{eqnarray}}

\noindent and we concentrate on the generalization of the Werner
state\cite{W}:

\begin{equation}
\rho_F=F \ket{\Phi^{+}}\bra{\Phi^{+}}+\frac{1-F}{3}\bigg(
\ket{\Phi^{-}}\bra{\Phi^{-}} + \ket{\Psi^{+}}\bra{\Psi^{+}} +
\ket{\Psi^{-}}\bra{\Psi^{-}}\bigg)
\end{equation}

\noindent This work concerns entanglement purification protocols.
At the beginning of these protocols, two persons, Alice and Bob,
share a large number of quantum states $\rho_F$, say
$\rho_F^{\otimes n}$, and they are allowed to communicate
classically, apply unitary transformations and perform projective
measurements. We place no restriction on the size of their ancilla
systems so that we lost no generality in restricting their local
operations to unitaries and projective measurements. In the end,
the quantum states $\Upsilon$ shared by Alice and Bob are to be a
close approximation of the maximally entangled states
$(\ket{\Phi^{+}} \bra{\Phi^{+}})^{\otimes m}$, or more precisely
we require the fidelity between $\Upsilon$ and $(\ket{\Phi^{+}}
\bra{\Phi^{+}})^{\otimes m}$ approaches one as $n$ goes to
infinity. We define the yield of such protocols to be $m/n$.

\subsection{Previous protocols}\label{1.2}

We briefly review some previous protocols in the following
subsections:

\subsubsection{Universal hashing}
Universal hashing was introduced in \cite{BDSW} and requires only
one-way classical communication. The scenario is the same as what
was described in section \ref{1.1}. The hashing method works by
having Alice and Bob each perform some local unitary operations on
the corresponding members of the shared bipartite quantum states.
They then locally measure some of the pairs to gain information
about the Bell states of the remaining unmeasured pairs. It was
shown that each measurement can be made to reveal almost 1 bit of
information about the unmeasured pairs; therefore, by measuring $n
S(\rho_F)$ pairs, Alice and Bob can figure out the identities of
the remaining unmeasured pairs with high probability. Once the
identities of the Bell states are known, Alice and Bob can convert
them into the standard states $\Phi^{+}$ easily. This protocol
distills a yield $D_H=[n - nS(\rho_F)]/n=1-S(\rho_F)$.

\subsubsection{The recurrence method}

\begin{figure}[h]
\centerline{\epsfig{file=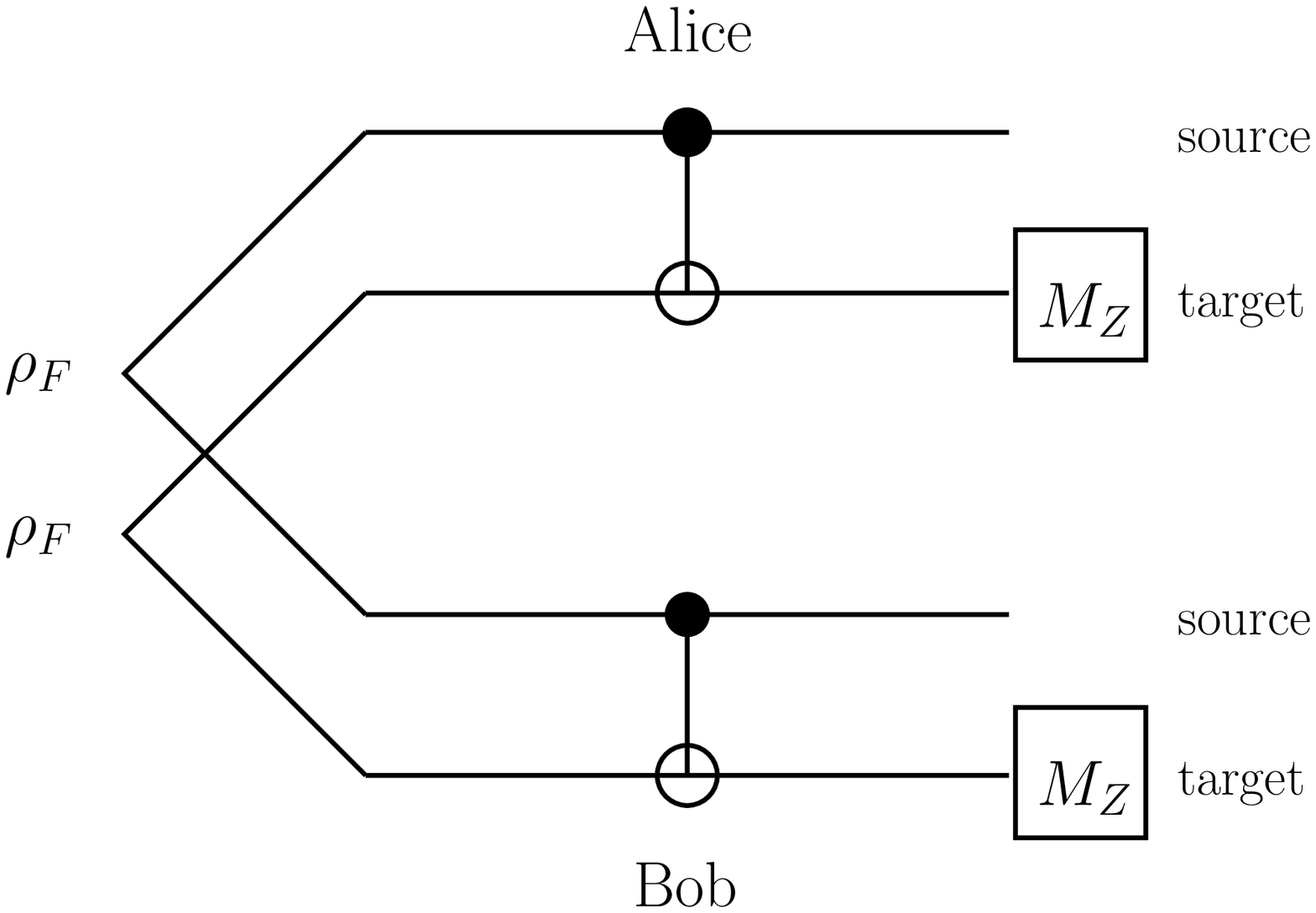, width=80mm}} 
\vspace*{13pt} \fcaption{\label{Recurrence}The recurrence method.}
\end{figure}

The recurrence method\cite{BDSW,BBPSSW} is illustrated in
fig.\ref{Recurrence}. Alice and Bob put the quantum states
$\rho_F^{\otimes n}$ into groups of two and apply XOR operations
to the corresponding members of the quantum states
$\rho_F^{\otimes 2}$, one as the source and one as the target.
They then take projective measurements on the target states along
the z-axis, and compare their measurement results with the side
classical communication channel. If they get identical results,
the source pair ``passed"; otherwise the source pair ``failed".
Alice and Bob then collect all the ``passed" pairs, and apply a
unilateral $\pi$ rotation $\sigma_x$ followed by a bilateral
$\pi/2$ rotation $B_x$\footnote{As mentioned in \cite{BDSW}, the
application of a $\sigma_x$ and $B_x$ rather than a twirl was
proposed by C. Macchiavello.}. This process is iterated until it
becomes more beneficial to pass on to the universal hashing. If we
denote the quantum states by
$\rho=p_{00}\ket{\Phi^{+}}\bra{\Phi^{+}}
+p_{01}\ket{\Psi^{+}}\bra{\Psi^{+}} +p_{10}\ket{\Phi^{-}}
\bra{\Phi^{-}}+p_{11}\ket{\Psi^{-}}\bra{\Psi^{-}}$, then this
protocol has the following recurrence relation:

{\setlength\arraycolsep{2pt}
\begin{eqnarray}
&p'_{00}=(p_{00}^2+p_{10}^2)/p_{pass};
& p'_{01}=(p_{01}^2 +p_{11}^2)/p_{pass}; \nonumber\\
&p'_{10}=2p_{01}p_{11}/p_{pass};
&p'_{11}=2p_{00}p_{10}/p_{pass};\nonumber
\end{eqnarray}}

\noindent and

\begin{equation}
p_{pass}=p_{00}^2+p_{01}^2+p_{10}^2+p_{11}^2+2p_{00}p_{10}+2p_{01}p_{11}\nonumber
\end{equation}

\subsubsection{The Maneva-Smolin method}
\begin{figure}[h]
\centerline{\epsfig{file=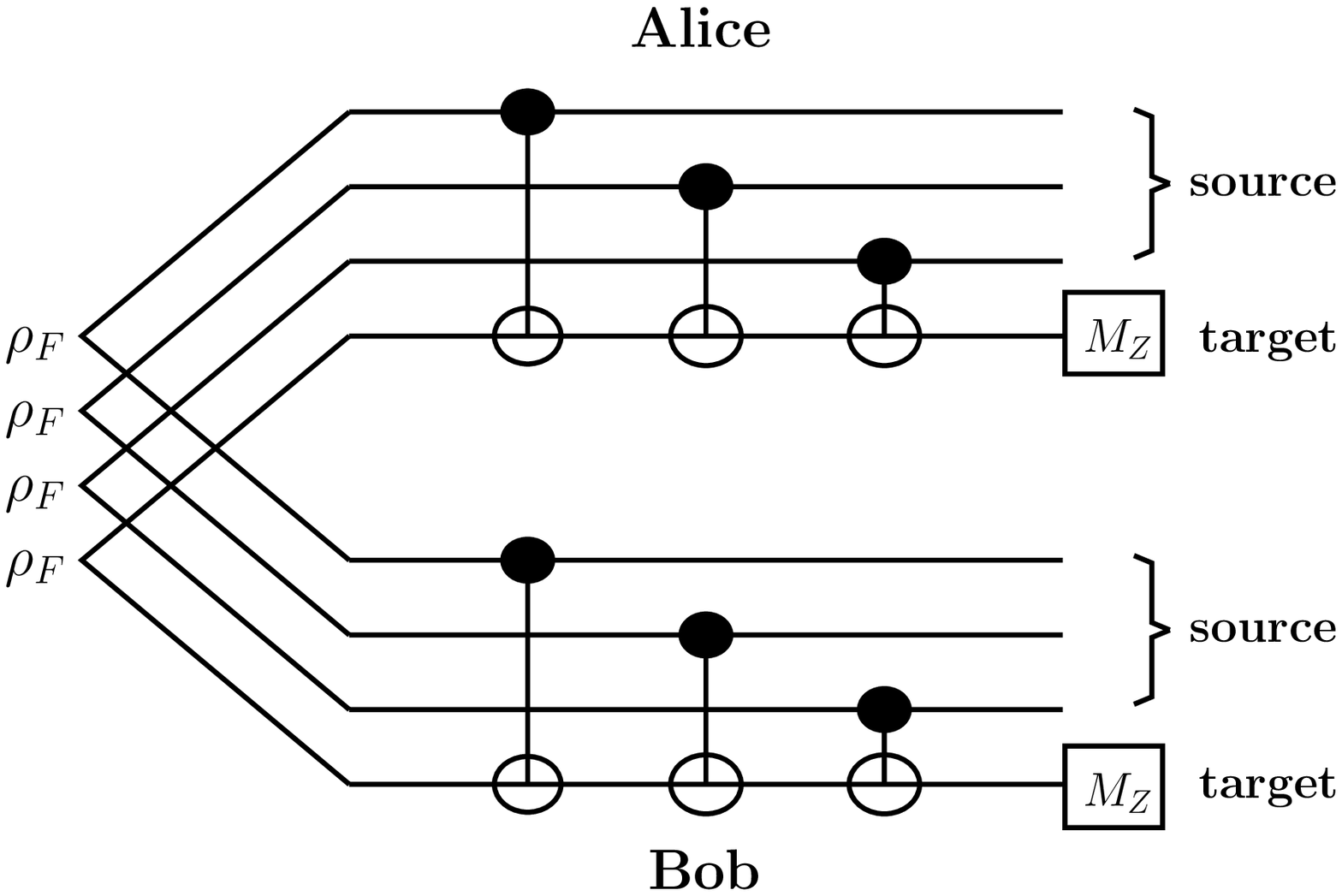, width=100mm}} 
\vspace*{13pt} \fcaption{\label{MS}The Maneva-Smolin method when
$m =4$.}
\end{figure}

The Maneva-Smolin method\cite{MS} is illustrated in fig.\ref{MS}.
Alice and Bob first choose a block size $m$ and put the quantum
states into groups of $m$. They then apply bipartite XOR gates
between each of the first $m-1$ pairs and the $mth$ pairs. After
that, they take measurements on these $mth$ pairs along the
z-axis, and compare their results with side classical
communication channel. If they get identical results, they perform
universal hashing on the corresponding $m-1$ remaining pairs; if
they get different results, they throw away all $m$ pairs. The
yield for this method is:

\begin{equation}
p_{pass}\frac{m-1}{m}\bigg(1-\frac{H(\textrm{passed source
states})}{m-1}\bigg) \nonumber
\end{equation}

\section{Entanglement purification protocol}
In section \ref{2.1}, we will present our new entanglement
purification protocol and compare its yield with the yields
obtained by the recurrence method \cite{BDSW} and the
Maneva-Smolin method \cite{MS}. In section \ref{2.2}, we will give
a closed-form expression for the yield of this new protocol.

\subsection{New protocol and improved yield}\label{2.1}

\begin{figure}[h]
\centerline{\epsfig{file=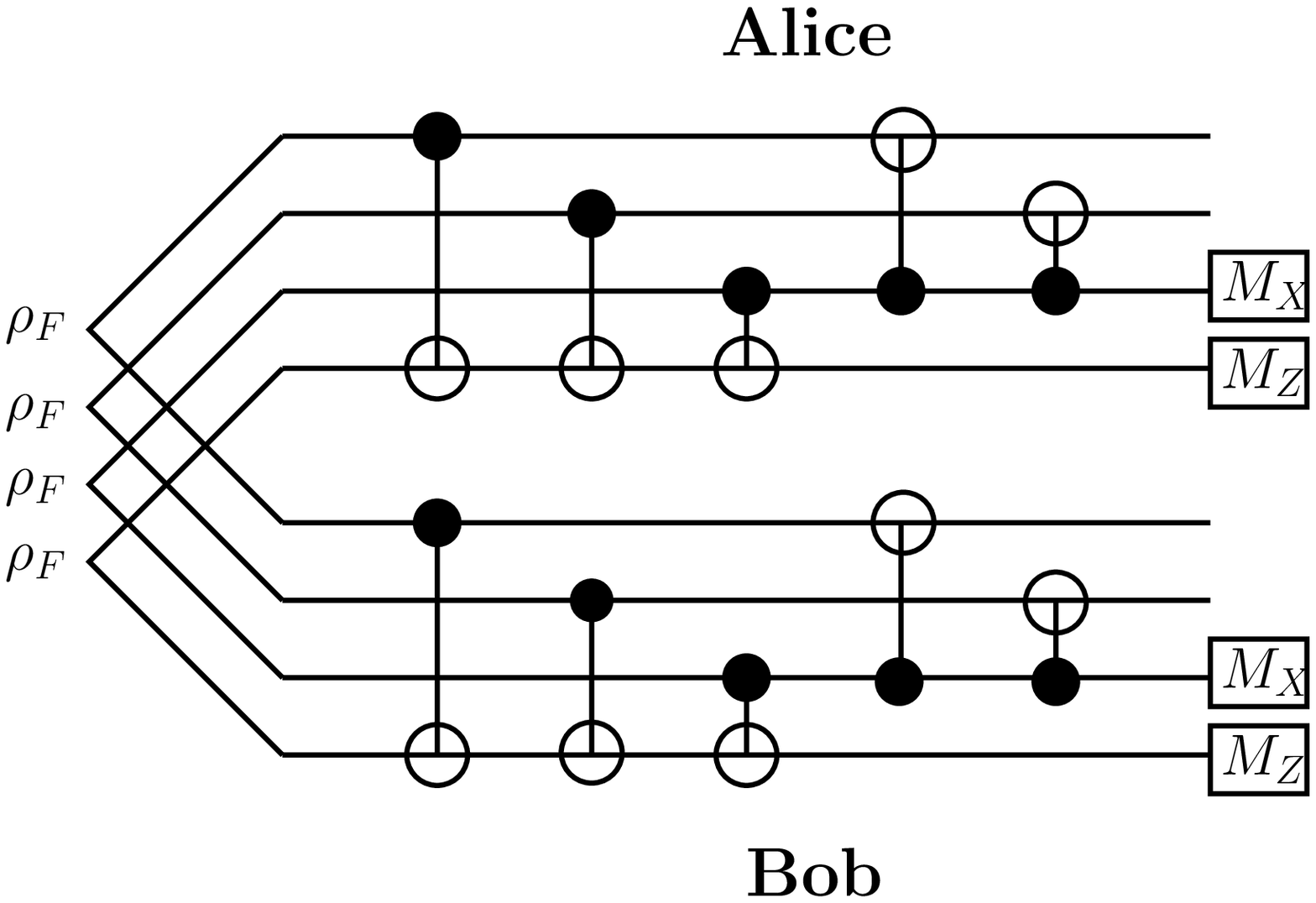, width=100mm}} 
\vspace*{13pt} \fcaption{\label{LS}Alice and Bob put the quantum
states $\rho_F^{\otimes n}$ into groups of four, and apply quantum
circuits consisting only of XOR gates. They then take measurements
on the third pair along the x-axis and the fourth pair along the
z-axis. With side classical communication, they compare their
results and if both results agree, they apply universal hashing on
the first two pairs. If either measurement result disagrees, they
throw away all four pairs. The yield of this protocol is plotted
on fig. \ref{yields}.}
\end{figure}

Our protocol is illustrated in fig.\ref{LS}. Alice and Bob share
the quantum states $\rho_F^{\otimes n}$ and put them into groups
of four. They then apply the quantum circuit shown in fig.\ref{LS}
and take measurements on the third and fourth pairs along the x-
and z-axis respectively. Using the side classical communication
channel, they can compare their results with each other. If they
get identical results on both measurements, they keep the first
and second pairs and apply universal hashing\cite{BDSW}. If either
of the two results disagrees, they throw away all four pairs.

The four pairs can be described by an 8-bit binary string, and
since these are mixed states they are in fact probability
distribution over all $256 (=2^8)$ possible 8-bit binary strings.
The quantum circuit consists only of XOR gates and therefore maps
the 8-bit binary strings, along with their underlying probability
distribution, bijectively to themselves. Let us call these
probability distributions $P(a_1 a_2 b_1 b_2 c_1 c_2 d_1 d_2)$ and
$P'(a_1 a_2 b_1 b_2 c_1 c_2 d_1 d_2)$.

\begin{figure}
\includegraphics[width=100mm]{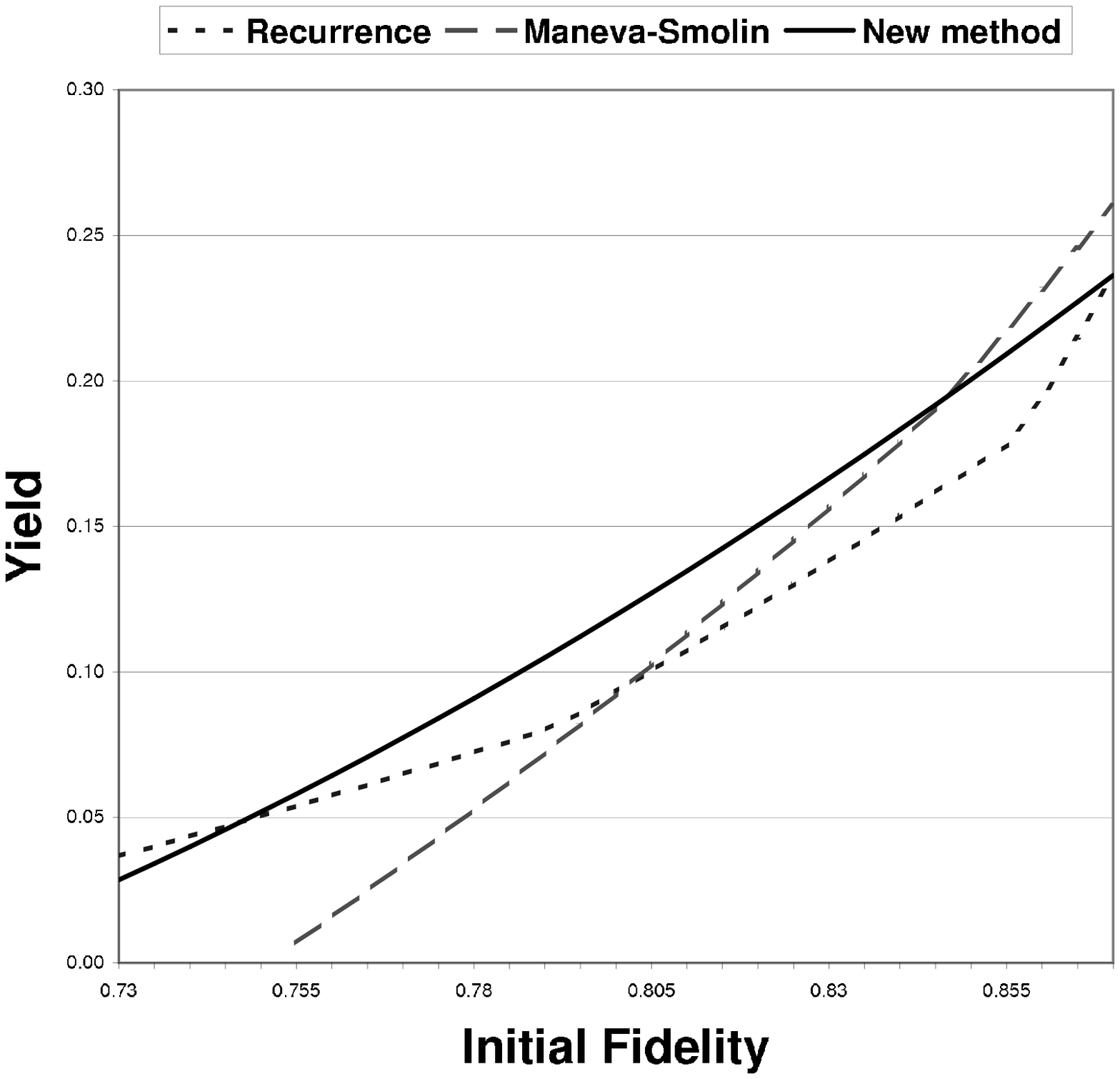}
\vspace*{13pt} \fcaption{\label{yields}The dotted line is the
yield for modified recurrence method \cite{BDSW}; the dash line is
for the Maneva-Smolin method \cite{MS}. The yield of our new
method is represented by the solid line, and there is an
improvement over the previous methods when the initial fidelity is
between $7.5$ and $8.45$.}
\end{figure}

Our quantum measurements on the third and fourth pairs are simply
looking at the 5th bit (measurement on the third pair along
x-axis) and the 8th bit (measurement on the fourth pair along
z-axis), where a ``0" means Alice and Bob getting identical
results and a ``1" means their getting opposite results. For
example, if the 8-bit binary string is
``$a_1a_2b_1b_2c_1c_2d_1d_2=00100111$", which corresponds to the
quantum states $\Phi^{+} \Phi^{-} \Psi^{+} \Psi^{-}$, then Alice
and Bob will get identical results on the third pair but opposite
results on the fourth. The ``pass" probability is
$p_{pass}=\sum_{a_1,a_2,b_1,b_2 c_2,d_1\in\{0,1\}}P'(a_1 a_2 b_1
b_2 0c_2d_10)$ and the post-measurement probability distribution
is $Q(a_1a_2b_1b_2)=\sum_{c_2,d_1\in\{0,1\}}P'(a_1 a_2 b_1 b_2
0c_2d_10)/p_{pass}$. The yield of this method\cite{MS} is:

\begin{equation}
\frac{p_{pass}}{2}\bigg(1-\frac{H(Q(a_1a_2b_1b_2))}{2}\bigg)
\label{closed-form1}
\end{equation}

\noindent where $H(Q(a_1a_2b_1b_2))$ is the Shannon entropy
function. Fig. \ref{yields} compares the yield of our new method
with the recurrence method and the Maneva-Smoline method.

\subsection{Closed-form expression}\label{2.2}

The quantum circuit that Alice and Bob apply to the quantum states
$\rho_F^{\otimes 4}$ consists only of XOR gates and therefore maps
the 8-bit binary strings bijectively to themselves. We call this
bijection $f$:

{\setlength\arraycolsep{2pt}
\begin{eqnarray}
f: \{0,1\}^8 & \longrightarrow & \{0,1\}^8 \nonumber\\
(a_1,a_2,b_1,b_2,c_1,c_2,d_1,d_2)&\longmapsto& (a_1\oplus d_1, a_2
\oplus c_2, b_1 \oplus d_1, b_2 \oplus c_2, \nonumber\\
& & \hspace*{15pt} a_1 \oplus b_1 \oplus c_1 \oplus d_1, c_2, d_1,
a_2 \oplus b_2 \oplus c_2 \oplus d_2)\nonumber
\end{eqnarray}}

\begin{table}[h]
\tcaption{\label{table}The quantum states that lead to identical
results for Alice and Bob.} \centerline{\footnotesize $G=(1-F)/3;
\Phi^{+}=00; \Psi^{+}=01; \Phi^{-}=10; \Psi^{-}=11;$}
\centerline{\footnotesize\smalllineskip
\begin{tabular}{c c c c}\\
\hline
$P(a_1a_2b_1b_2c_1c_2d_1d_2)$ & $a_1a_2b_1b_2c_1c_2d_1d_2$ & $f(a_1a_2b_1b_2c_1c_2d_1d_2)$ & $\textit{tr}_{c,d}\big(f(a_1a_2b_1b_2c_1c_2d_1d_2)\big)$ \\
\hline
$F^4$ &00000000 &00000000 &0000 \\
$G^4$ &01010101 &00000100 &0000 \\
$G^4$ &10101010 &00000010 &0000 \\
$G^4$ &11111111 &00000110 &0000 \\
\hline
$F^2G^2$ &00010001 &00010000 &0001 \\
$F^2G^2$ &01000100 &00010100 &0001 \\
$G^4$ &10111011 &00010010 &0001 \\
$G^4$ &11101110 &00010110 &0001 \\
\hline
$F^2G^2$ &00101000 &00100000 &0010 \\
$G^4$ &01111101 &00100100 &0010 \\
$F^2G^2$ &10000010 &00100010 &0010 \\
$G^4$ &11010111 &00100110 &0010 \\
\hline
$FG^3$ &00111001 &00110000 &0011 \\
$FG^3$ &01101100 &00110100 &0011 \\
$FG^3$ &10010011 &00110010 &0011 \\
$FG^3$ &11000110 &00110110 &0011 \\
\hline
$F^2G^2$ &00010100 &01000100 &0100 \\
$F^2G^2$ &01000001 &01000000 &0100 \\
$G^4$ &10111110 &01000110 &0100 \\
$G^4$ &11101011 &01000010 &0100 \\
\hline
$F^2G^2$ &00000101 &01010100 &0101 \\
$F^2G^2$ &01010000 &01010000 &0101 \\
$G^4$ &10101111 &01010110 &0101 \\
$G^4$ &11111010 &01010010 &0101 \\
\hline
$F^2G^2$ &00111100 &01100100 &0110 \\
$G^4$ &01101001 &01100000 &0110 \\
$G^4$ &10010110 &01100110 &0110 \\
$F^2G^2$ &11000011 &01100010 &0110 \\
\hline
$FG^3$ &00101101 &01110100 &0111 \\
$FG^3$ &01111000 &01110000 &0111 \\
$FG^3$ &10000111 &01110110 &0111 \\
$FG^3$ &11010010 &01110010 &0111 \\
\hline
$F^2G^2$ &00100010 &10000010 &1000 \\
$G^4$ &01110111 &10000110 &1000 \\
$F^2G^2$ &10001000 &10000000 &1000 \\
$G^4$ &11011101 &10000100 &1000 \\
\hline
$F^2G^2$ &00110011 &10010010 &1001 \\
$G^4$ &01100110 &10010110 &1001 \\
$G^4$ &10011001 &10010000 &1001 \\
$F^2G^2$ &11001100 &10010100 &1001 \\
\hline
$F^2G^2$ &00001010 &10100010 &1010 \\
$G^4$ &01011111 &10100110 &1010 \\
$F^2G^2$ &10100000 &10100000 &1010 \\
$G^4$ &11110101 &10100100 &1010 \\
\hline
$FG^3$ &00011011 &10110010 &1011 \\
$FG^3$ &01001110 &10110110 &1011 \\
$FG^3$ &10110001 &10110000 &1011 \\
$FG^3$ &11100100 &10110100 &1011 \\
\hline
$FG^3$ &00110110 &11000110 &1100 \\
$FG^3$ &01100011 &11000010 &1100 \\
$FG^3$ &10011100 &11000100 &1100 \\
$FG^3$ &11001001 &11000000 &1100 \\
\hline
$FG^3$ &00100111 &11010110 &1101 \\
$FG^3$ &01110010 &11010010 &1101 \\
$FG^3$ &10001101 &11010100 &1101 \\
$FG^3$ &11011000 &11010000 &1101 \\
\hline
$FG^3$ &00011110 &11100110 &1110 \\
$FG^3$ &01001011 &11100010 &1110 \\
$FG^3$ &10110100 &11100100 &1110 \\
$FG^3$ &11100001 &11100000 &1110 \\
\hline
$F^2G^2$ &00001111 &11110110 &1111 \\
$G^4$ &01011010 &11110010 &1111 \\
$G^4$ &10100101 &11110100 &1111 \\
$F^2G^2$ &11110000 &11110000 &1111 \\
\hline
\end{tabular}}
\end{table}

\noindent In table \ref{table}, we list the quantum states that
lead to identical measurement results for Alice and Bob and their
probabilities in the ensemble $\rho_F^{\otimes 4}$. Therefore, we
can write down expressions for the terms $p_{pass}$ and $H(Q(a_1
a_2 b_1 b_2)$ in equation(\ref{closed-form1}) as follows:

{\setlength\arraycolsep{2pt}
\begin{eqnarray}
p_{pass}&=&F^4+18F^2G^2+24FG^3+21G^4 \label{closed-form2} \\
H(Q(a_1a_2b_1b_2))&=&-\Big(\frac{F^4+3G^4}{p_{pass}}\Big)\log_2\Big(\frac{F^4+3G^4}{p_{pass}}\Big)
-9\Big(\frac{2F^2G^2+2G^4}{p_{pass}}\Big)\log_2\Big(\frac{2F^2G^2+2G^4}{p_{pass}}\Big)\nonumber\\
&&-6\Big(\frac{4FG^3}{p_{pass}}\Big)\log_2\Big(\frac{4FG^3}{p_{pass}}\Big)\label{closed-form3}
\end{eqnarray}}

\noindent where $G=(1-F)/3$.

\section{Conclusions}
We presented a new protocol for entanglement purification assisted
by two-way classical communication. It was shown in \cite{BDSW}
that such a protocol corresponds to quantum capacity assisted by
two-way classical communication of the quantum depolarizing
channel, and hence we have a new lower bound for this capacity. In
section \ref{2.1}, we applied the new protocol to the quantum
states

$$\rho_F=F \ket{\Phi^{+}}\bra{\Phi^{+}}+\frac{1-F}{3}\bigg(
\ket{\Phi^{-}}\bra{\Phi^{-}} + \ket{\Psi^{+}}\bra{\Psi^{+}} +
\ket{\Psi^{-}}\bra{\Psi^{-}}\bigg);$$

\noindent however, our method also works for any Bell-diagonal
quantum states

$$\rho= p_{00} \ket{\Phi^{+}}\bra{\Phi^{+}}
+p_{01}\ket{\Psi^{+}}\bra{\Psi^{+}}+p_{10} \ket{\Phi^{-}}
\bra{\Phi^{-}} +p_{11}\ket{\Psi^{-}}\bra{\Psi^{-}}.$$

\noindent Equation (\ref{closed-form2}) and (\ref{closed-form3})
then become

{\setlength\arraycolsep{2pt}
\begin{eqnarray}
p_{pass}&=& \big(p_{00}^4+p_{01}^4+p_{10}^4+p_{11}^4\big) + 6
\times 4 p_{00}p_{01}p_{10}p_{11}+ 3 \times \sum_{\substack{i,j\in\{0,1\}^2 \\
i\neq j} } 2 p_i^2p_j^2  \nonumber \\
H(Q(a_1a_2b_1b_2))&=&-\Big(\frac{p_{00}^4+p_{01}^4+p_{10}^4+p_{11}^4}{p_{pass}}\Big)
\log_2\Big(\frac{p_{00}^4+p_{01}^4+p_{10}^4+p_{11}^4}{p_{pass}}\Big)\nonumber\\
&& - 6 \times \Big(\frac{4
p_{00}p_{01}p_{10}p_{11}}{p_{pass}}\Big)\log_2\Big(\frac{4
p_{00}p_{01}p_{10}p_{11}}{p_{pass}}\Big) \nonumber\\
&&-3\times\Big(\frac{2p_{00}^2p_{01}^2+2p_{10}^2p_{11}^2}{p_{pass}}\Big)\log_2\Big(\frac{2p_{00}^2p_{01}^2+2p_{10}^2p_{11}^2}{p_{pass}}\Big)\nonumber\\
&&-3\times\Big(\frac{2p_{00}^2p_{10}^2+2p_{01}^2p_{11}^2}{p_{pass}}\Big)\log_2\Big(\frac{2p_{00}^2p_{10}^2+2p_{01}^2p_{11}^2}{p_{pass}}\Big)\nonumber\\
&&-3\times\Big(\frac{2p_{00}^2p_{11}^2+2p_{01}^2p_{10}^2}{p_{pass}}\Big)\log_2\Big(\frac{2p_{00}^2p_{11}^2+2p_{01}^2p_{10}^2}{p_{pass}}\Big)\nonumber
\end{eqnarray}}

\noindent With these equations, we can combine the recurrence
method and our new method: we start with the recurrence method and
pass on to our new method rather than universal hashing. Indeed,
there are improvements, but they occur over segments of narrow
regions and the improvements are insignificant. Therefore we
believe these improvements have only to do with the number of
recurrence steps performed before passing on to universal hashing,
and we will spare the readers with the details.

An obvious direction is to look for new protocols with better
yields. In particular, as was raised by E.N. Maneva and J.A.
Smolin in \cite{MS}, is there a way to iterate this protocol
rather than passing on immediately to universal hashing? It is
conceivable that, with such an adaption, we can have improvements
over a wider range of initial fidelity.

After the completion of this work, it came to our attention
similar works have been carried out in \cite{VV,HDM}.

\nonumsection{References}


\begin{thebibliography}{000}

\bibitem{NC}
M.A. Nielsen and I.L. Chuang, {\it Quantum computation and quantum
information}, Cambridge University Press (2000)

\bibitem{BBCJPW}
C.H. Bennett, G. Brassard, C. Cr\'{e}peau, R. Jozsa, A. Peres and
W. Wootters, {\it Teleporting an unknown quantum state via dual
classical and EPR channels}, Phys. Rev. Lett., 70, pp. 1895-1899
(1993)

\bibitem{BW}
C.H. Bennett and S.J. Wiesner, {\it Communication via one- and
two-particale operators on Einstein-Podolsky-Rsen states}, Phys.
Rev. Lett., 69, pp. 2881-2884 (1992)

\bibitem{S}
P.W. Shor, {\it Polynomial-Time Algorithms for Prime Factorization
and Discrete Logarithms on a Quantum Computer}, SIAM J. Comput.,
26, pp. 1484-1509 (1997)

\bibitem{G}
L. Grover, {\it A fast quantum mechanical algorithm for database
search}, Proceddings of the 28th Annual ACM Symposium on Theory of
Computing, pp. 212-219 (1996)

\bibitem{BDSW}
C.H. Bennett, D.P. DiVincenzo, J.A. Smolin and W.K. Wootters, {\it
Mixed State Entanglement and Quantum Error Correction}, Phys. Rev.
A, 54, pp. 3824-3851 (1996), quant-ph/9604024.

\bibitem{MS}
E.N. Maneva and J.A. Smolin, {\it Improved two-party and
multi-party purification protocols}, quant-ph/0003099.

\bibitem{W}
R.F. Werner, {\it Quantum states with Einstein-Podolsky-Rosen
correlations admitting a hidden-variable model}, Phys. Rev. A, 40,
pp. 4277-4281 (1989)

\bibitem{BBPSSW}
C.H. Bennett, G. Brassard, S. Popescu, B. Schumacher, J.A. Smolin
and W.K. Wooters, {\it Purification of Noisy Entanglement and
Faithful Teleportation via Noisy Channels}, Phys. Rev. Lett., 76,
pp. 722-725 (1996)

\bibitem{VV}
K.G.H. Volbrecht and Frank Verstraete, {\it Interpolation of
recurrence and hashing entanglement distillation protocols},
quant-ph/0404111

\bibitem{HDM}
E. Hostens, J. Dehaene and B.D. Moor, {\it Asymptotic adaptive
bipartite entanglement distillation protocol}, quant-ph/0602205

\end{thebibliography}
\end{document}